\documentclass[twocolumn,amsmath,amssymb,showpacs,superscriptaddress,prl]{revtex4-1}

\usepackage{appendix}
\usepackage{graphicx}
\usepackage{braket}
\usepackage{color} 
\usepackage{bbm}

\begin{document}

\title{Surface-state spin textures in strained bulk HgTe: \\
strain-induced topological phase transitions
 }

\author{Frank Kirtschig}
\affiliation{Institute for Theoretical Solid State Physics, IFW Dresden, PF 270116, 01171 Dresden, Germany}
\author{Jeroen van den Brink}
\affiliation{Institute for Theoretical Solid State Physics, IFW Dresden, PF 270116, 01171 Dresden, Germany}
\affiliation{Department of Physics, Dresden University of Technology, 01062 Dresden, Germany}
\affiliation{Department of Physics, Harvard University, Cambridge, Massachusetts 02138, USA}
\author{Carmine Ortix}
\affiliation{Institute for Theoretical Solid State Physics, IFW Dresden, PF 270116, 01171 Dresden, Germany}
\affiliation{Institute for Theoretical Physics, Center for Extreme Matter and Emergent Phenomena, Utrecht University, Leuvenlaan 4, 3584 CE Utrecht, Netherlands}
\date{\today}

\begin{abstract}
The opening of a band gap due to compressive uniaxial strain renders bulk HgTe a strong three-dimensional topological insulators with protected gapless surface states at any surface. By employing a six-band ${\bf k \cdot p}$ model, we determine the spin textures of the topological surface states of strained HgTe using their close relations with  the mirror Chern numbers of the system and the orbital composition of the surface states. We show that at surfaces with ${\cal C}_{2 v}$ point group symmetry 
an increase in the strain magnitude triggers a topological phase transition
where the winding number of the surface state spin texture is flipped while the four topological invariants characterizing the bulk band structure of the material are unchanged.  
\end{abstract}


\maketitle

\begin{figure}
\includegraphics[width=.9\columnwidth]{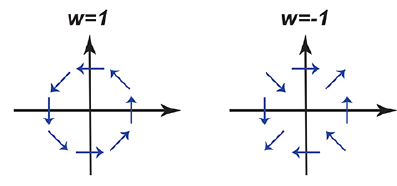}
\caption{(color online) Sketch of the possible spin textures of Dirac-like surface states in strong 3D TI. The left panel corresponds to a right-handed helical structures with winding number $w=1$. In the right panel $w=-1$.}
\label{fig:fig0}
\end{figure}

\paragraph{Introduction -- } 
Topological insulators (TIs) are new quantum states of matter whose theoretical prediction and experimental verification has had a tremendous impact in the field of fundamental condensed matter physics \cite{kan05,kan05b,fu06,ber06,wu06,kon07,fu07,fu07b,moo07,zha09,hsi08,xia09,che09,hsi09,ras13,has10,qi11}, and for potential applications in spintronics and quantum computation \cite{akh09}.   
Time-reversal (TR) invariant TIs are insulating in the bulk but they do possess gapless surface states topologically protected by TR symmetry \cite{kan05,fu06}. These metallic surface states are spin-momentum locked: surface electrons with opposite spins counterpropagate at the sample boundaries.  For three-dimensional (3D)  TIs, Bi$_2$Se$_3$ \cite{zha09}, Bi$_{14}$Rh$_3$I$_9$ \cite{ras13}, and $\beta$-HgS \cite{vir11,vir13} to name but a few, the existence of these topological surface states (TSS) can be directly inferred from the four $\mathbb{Z}_2$ indices characterizing the bulk band structure of a 3D TR invariant insulator \cite{fu07,fu07b,moo07}. However, both the strong  $\nu_{0}$ index and the three weak $\left\{\nu_1,\nu_2,\nu_3 \right\}$ indices make no assertion on the nature of the spin textures of the  surface states which, realizing a vortex structure in momentum space, can be characterized topologically by the winding number (the topological charge of the vortex)  of the planar unit spin $(n_i, n_j)=(S_i, S_j) / \sqrt{S_i^2+S_j^2}$. It is  defined by 
$$w=\oint_{\mathbb{C}} \dfrac{d {\bf k}}{2\pi} \cdot \left[n_i \nabla_{\bf k} n_j - n_j \nabla_{\bf k} n_i  \right],$$
where $\mathbb{C}$ is a closed loop in momentum space encircling the essential degeneracy point of the topological surface state, guaranteed by TR invariance. 
Generally speaking, the two winding numbers $w=\pm 1$ [c.f. Fig.~\ref{fig:fig0}] are equally compatible for linear Dirac cones, and the specific value is independent of the $\mathbb{Z}_2$ topological indices of the bulk band structure. In many strong 3D TIs with a single Dirac cone on the surface, however, additional point group symmetries at the surfaces pin the spin texture winding number to $w=1$. This occurs at the high ${\cal C}_{3 v}$ \cite{fu09,liu10} symmetry surfaces of materials with a rhombohedral crystal structure such as  Bi$_2$Se$_3$, or at the  ${\cal C}_{4 v}$ point group symmetric surfaces \cite{ort14} of cubic materials such as $\beta$-HgS \cite{note}. 
For surfaces where the symmetry is lowered, a similar assertion cannot be made. 

The aim of this Letter is to show that for ${\mathcal C}_{2v}$ point group symmetric surfaces, the surface state spin textures of compressively strained bulk HgTe -- a strong 3D TI whose non-trivial topological properties  have been experimentally verified by quantum Hall measurements \cite{bru11} -- have a topological charge that can be flipped from $w=1$ to $w=-1$ by continuously increasing the strain magnitude. 
We will use the close relations \cite{leg15} connecting the winding number of the spin textures, the mirror Chern numbers of the system and the orbital composition of the topological surface states, and thereby demonstrate, within a six-band  ${\bf k \cdot p}$ Kane model, that for a uniaxial strain along the $(100)$ direction, the orbital character of the topological surface states at the $(010)$ and $(001)$  surfaces depends sensitively upon the strain magnitude. This ultimately leads to a change in the nature of the surface state spin textures at a critical strain magnitude.

\paragraph{Topological Surface State Dirac points -- } Pristine HgTe is a zero gap semiconductor with the Fermi energy in the middle of the fourfold degenerate light-hole (LH) and heavy-hole (HH) $\Gamma_8$ states at the BZ center \cite{fu07,chu11}.  The topological nature of the electronic states in this material cannot be inferred from these $p_{3/2}$ atomic levels \cite{fu07} but rather follows from the inverted band ordering at the zone center of the LH $\Gamma_8$ band, which is particle-like, and the $\Gamma_6$ $s$-band, which is hole-like. In normal semiconductors, such as CdTe, the $\Gamma_6$ band forms the conduction band while the LH $\Gamma_8$ band represents one of the valence bands. 
This inverted band ordering, which is an immediate consequence of the strong spin-orbit coupling of Hg, 
establishes this material to be topologically non-trivial since two bands of opposite parities have level crossed with respect to the normal band ordering. 
By externally applying a compressive uniaxial strain, the fourfold degeneracy of the $\Gamma_8$ states at the zone center is lifted and thus a gap at the Fermi energy opens up \cite{fu07,bru11}. In addition, the parity eigenvalues of the occupied bands are unchanged, which thereby establishes compressively strained HgTe as a strong 3D TI. 

The bulk-boundary correspondence \cite{has10,qi11} then guarantees the existence of TSS with a conical dispersion at any surface, and the surface Kramer's doublet -- the Dirac point -- sitting at the surface BZ center. To verify this, we rely on an effective low-energy theory based upon a ${\bf k \cdot p}$ expansion of the lowest energy bands around the $\Gamma$ point of the BZ. 
This approach has successfully described the Quantum Spin Hall effect in HgTe/CdTe quantum wells \cite{ber06,kon07}. We thus employ the six-band Kane model for the $\Gamma_{6,8}$ bands \cite{win03}, where the influence of a compressive uniaxial strain, which, without loss of generality, we assume along the $\hat{x}$ direction, is taken into account via the Bir-Pikus Hamiltonian [see the Supplemental Material]. 
The Luttinger and ${\bf k \cdot p}$ parameters are based on the $T=0$ band structure of pristine HgTe \cite{nov05}. 
We can establish the presence and the electronic characteristic of the surface Kramer's doublet at the $(100)$, $(010)$ and $(001)$ surfaces by solving the ${\bf k \cdot p}$ model at the surface BZ center in the half-infinite space $x>0$, $y>0$, and $z>0$ respectively, using the general method outlined in Ref.~\onlinecite{ort14}. 

\begin{figure}
\includegraphics[width=.95\columnwidth]{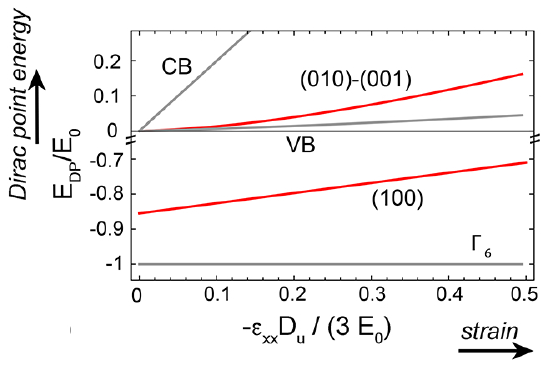}
\caption{(color online) Behavior of the surface Dirac point in compressively strained HgTe as a function of the strain magnitude $\epsilon_{xx}$ at the $(100)$,$(010)$, and $(001)$ surfaces (red lines). We also show the conduction and valence band edges as well as the $\Gamma_6$ band edge (gray lines). At the $(100)$ surface, the surface Dirac point is buried within the HH valence band but resides in the bandgap of the LH-$\Gamma_6$ TI bulk.}
\label{fig:fig1}
\end{figure}

Fig.~\ref{fig:fig1}(a) shows the behavior of the surface Dirac point energy $E_{DP}$ as a function of the uniaxial strain magnitude $\epsilon_{xx}$ renormalized by the factor $D_u / ( 3 E_0)$ where $E_0$ is the distance of the $\Gamma_6$ band edge from the direct BZ center midgap
 while $D_u$ is the deformation potential of HgTe. 
At the $(100)$ surface, the surface Kramer's doublet is buried within the HH valence band, while on the two other  surfaces  it resides in the indirect bulk gap of the system. This termination dependence is also reflected in the behavior of the penetration depth of the surface states [see the Supplemental Material]. Specifically, the surface states at the $(010)$ and $(001)$ surfaces are characterized by a diverging decay length in the $\epsilon_{xx} \rightarrow 0$ limit, which implies that at these planes the TSS penetrate more deeply into the bulk as compared to the $(100)$ TSS. 

These different electronic characteristics can be attributed to the different nature of the Dirac wavefunction of the surface BZ center. At the $(100)$ plane, indeed, the surface state Dirac wavefunction is all made of LH and $\Gamma_6$ states. A uniaxial strain along the $\hat{x}$ direction preserves the axial rotation symmetry in the plane, and thus at the surface $\Gamma$ point with momentum $k_y \equiv k_z \equiv 0$ the total angular momentum $J_x$ is a good quantum number \cite{ber06,ort14}. This, in turn, implies the absence of any mixing between the $\ket{J=3/2 ; J_x=\pm 3/2}$ HH states and the $J_x = \pm 1/2$ LH and $\Gamma_6$ states. Henceforth, the HH bands play the role of inserted ``parasitic" bands \cite{ort14} on top of the LH-$\Gamma_6$ TI bulk, in the bandgap of which the surface Dirac point resides [c.f. Fig.~\ref{fig:fig1}]. 
This does not hold true at the $(010)$ and $(001)$ planes where the uniaxial strain along the $\hat{x}$ direction breaks the in-plane rotation symmetry, thereby leading to an effective hybridization between the $J_{y,z}=\pm 1/2$ states with $J_{y,z}=\pm 3/2$ HH states. The surface Dirac wavefunction becoming a superposition of $\Gamma_{6,8}$ localized states is then pushed out of the HH bulk bandwidth \cite{ber10} and remerges in the full bandgap of the system, in agreement with the features encountered in the Fano model \cite{fan61}.

\begin{figure}
\includegraphics[width=.9\columnwidth]{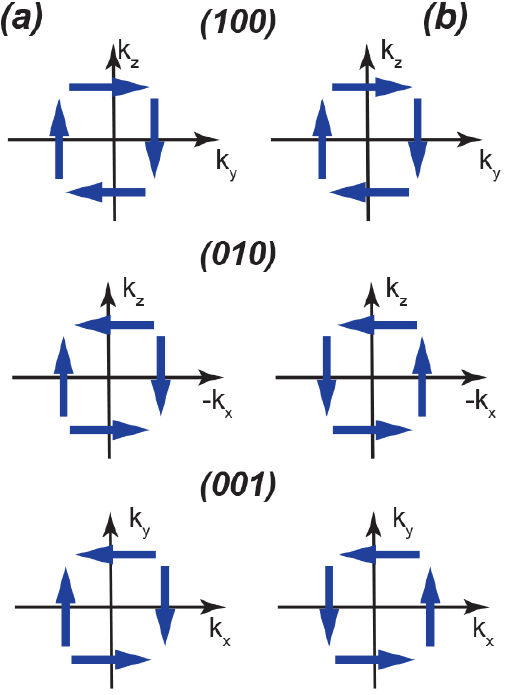}
\caption{(color online) (a) Pseudospin textures for the TSS of HgTe in presence of a uniaxial strain along the $(100)$ direction at the $(100)$ (top panels), $(010)$ (middle panels) and $(001)$ (bottom panels) crystal planes. (b) Physical spin textures for strain smaller than the critical one $-\epsilon_{xx} < -\epsilon_c$. For larger strain the physical spin textures corresponds to the pseudospin ones.}
\label{fig:fig2}
\end{figure}

\paragraph{Mirror Chern numbers and pseudospin textures --} To proceed further, we now introduce the notion of  mirror Chern numbers (MCN). The MCN are topological invariants, which are protected by mirror symmetries. In the absence of strain and neglecting the bulk inversion asymmetry of the zincblende crystal structure, HgTe has nine mirror planes and correspondingly the Kane model Hamiltonian is invariant under these symmetry operations. The presence of an uniaxial strain along the $(100)$ direction  reduces the number of mirror planes but preserves the mirror symmetry with respect to the  $(100)$, $(010)$, and $(001)$ planes. 
Since the Kane model Hamiltonian commutes with the corresponding mirror symmetry operations at the three planes $k_{x,y,z} \equiv 0$, all eigenstates can be classified according to their $\pm i $ mirror parity. This allows to define two time-reversal related Chern numbers $\mathcal{C}_{\pm i}$  whose sum vanishes but with a difference $n_{\cal M}=({\mathcal C}_{i} - {\mathcal C}_{-i})/2$, which is an integer $\mathbb{Z}$ topological invariant and defines the MCN.

We have computed the MCNs of the full six-band Kane model Hamiltonian using its decomposition at the mirror planes in terms of the nine Gell-Mann matrices, and subsequently employed the elegant formulation of Ref.~\onlinecite{bar12} to derive the Chern numbers for the corresponding continuum models. This allows us to avoid an effective  two-band modelling which can only be introduced {\it ad hoc}.  We find  a MCN $n_{\cal M} \equiv -1$ at the $k_{y,z} \equiv 0$ planes, whereas $n_{\cal M} \equiv 1$ at the $k_x \equiv 0$ plane. And indeed, under proper coordinate transformations, the continuum ${\bf k \cdot p}$ Hamiltonians at the $k_{y,z} \equiv 0$ planes cannot be adiabatically transformed into the $k_x \equiv 0$ one without closing the bulk band gap. The three  MCNs indicated above  allow us to immediately derive a \emph{pseudospin} texture for the TSS at the $(100)$, $(010)$ and $(001)$ planes, as explained below. At the $(100)$ surface, the projection of the two unbroken mirror planes $(010)$, $(001)$ define two mirror invariant lines where the TSS  can be classified according to their mirror eigenvalues. 
We can thus define a pseudospin vector  $\boldsymbol{\sigma}$ with components related  to the mirror operators ${\cal M}_{y,z}$ by $\sigma_{y,z}=-  i {\cal M}_{y,z} $. The dispersion of the TSS can be then written in terms of this pseudospin as 
\begin{equation} 
H_{eff}^{(100)} =  v_F^z k_y \sigma_z - v_F^y k_z \sigma_y, 
\end{equation}
where the sign of the two Fermi velocities $v_F^{z,x}$ is uniquely determined by the MCNs of the system via the bulk-edge correspondence for the mirror invariant planes $k_{y,z} \equiv 0$. Specifically we have $\text{sgn}(v_F^{z,y}) = n_{\cal M}^{k_{z,y}=0}$ which yields the pseudospin texture shown in the top panel of Fig.~\ref{fig:fig2}(a).  It exhibits an helical structure with a left-handed helicity for the surface state conduction band, and a right-handed one for the valence band, in perfect agreement with density functional theory studies \cite{wan15,rau15}. 
At the opposite $(\bar{1} 0 0 )$ surface the sign of both two Fermi velocities are flipped, which changes the helicity of the pseudospin texture but still preserves the pseudospin texture winding number $w=\text{sgn} (v_F^z \times v_F^y)=1$. A similar analysis at the $(010)$ and $(001)$ surface terminations yields the pseudospin textures shown in Fig.~\ref{fig:fig2}(a). At these surfaces the different values of the two MCN for the mirror invariant planes yield a pseudospin texture with an opposite winding number $w=-1$.

\begin{figure}
\includegraphics[width=.87\columnwidth]{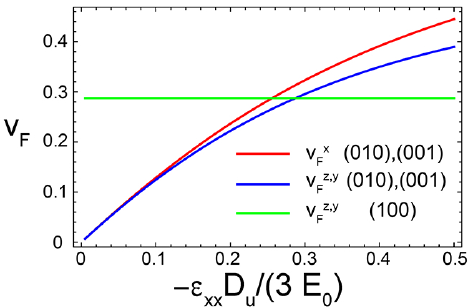}
\caption{(color online) Behavior of the magnitudes of the Fermi velocities $|v_F|$ in the TSS of strained HgTe as a function of the strain magnitude $\epsilon_{xx}$. The TSS at the $(010)$ and $(001)$ planes exhibit an anisotropic behavior since the uniaxial strain along the $\hat{x}$ direction  lower the point group symmetry at that surfaces to $\mathcal{C}_{2v}$.}
\label{fig:fig3}
\end{figure} 

Fig.~\ref{fig:fig3} shows the behavior of the magnitude of the TSS Fermi velocities $|v_F|$ as a function of the strain magnitude $\epsilon_{xx}$.  We find that for the TSS at the $(100)$ plane, the two Fermi velocities $v_F^{y,z}$  have equal magnitudes. Therefore, the TSS display a global $\mathcal{U}(1)$ rotational symmetry similarly to the case of for instance Bi$_2$Se$_3$ \cite{fu09}. As a two-dimensional $\mathbf{k \cdot p}$ for the pseudospin one-half surface Kramer doublet explicitly shows \cite{ort14}, this is an immediate consequence of the fourfold rotational symmetry along the $\hat{x}$ axis. And indeed, at the  $(010)$ and $(001)$ planes where the uniaxial strain along the $(001)$ direction lowers the surface point group symmetry from $\mathcal{C}_{4v}$ to $\mathcal{C}_{2v}$, we find the Fermi velocities of the TSS to differ, with an anisotropy that is enhanced by increasing the strain magnitude.

\paragraph{Spin textures --} The pseudospin textures of the TSS have a close relation to the physical spin textures. This can be found introducing the projector operator  $\mathcal{P}_{x,y,z}=\left(\mathbbm{1} - i \mathcal{M}_{x,y,z}  \right) / 2$ onto the subspaces of the Kane model Hamiltonian with mirror parity $+i$, and considering their effect on the physical spin operators $\mathcal{S}_{x,y,z}$ at the mirror planes. It can be shown that $\mathcal{P}_i \mathcal{S}_j \mathcal{P}_i \equiv 0$ for $i \neq j$, which simply states that at the surface mirror invariant lines, the physical spin can be either parallel or antiparallel to the pseudospin.  For SmB$_6$, the physical spin was found to be always parallel to the pseudospin, and thus the knowledge of the MCNs provides us a robust classification of the topological surface state spin textures \cite{leg15}. This, however, does not hold true for HgTe. The projected spin operator of the $\Gamma_{6,8}$ bands has both positive and negative eigenvalues, and thus the relation between the pseudospin and the physical spin depends on the orbital composition of the surface states. By evaluating the projected spin operator eigenvalues of the TSS along the mirror invariant lines, we find that the physical spin is always parallel to the pseudospin except at the $k_z \equiv 0$ and $k_y \equiv 0$ mirror invariant lines for the $(010)$ and $(001)$ surface terminations respectively. At these mirror invariant lines, the physical spin is indeed antiparallel to the pseudospin below a critical strain magnitude 
$|\epsilon_c|$, and parallel above it. This implies that contrary to the TSS pseudospin textures at the $(010)$ and $(001)$ surface terminations, which are characterized by a $w=-1$ winding number independent of the strain magnitude, the physical spin textures exhibit a right-handed $w=1$  helical structure for small strain [c.f. Fig.~\ref{fig:fig3}(b)]. The $w=1$ spin textures equal to the pseudospin textures are then restored for larger strain values.  
\begin{figure}
\includegraphics[width=.9\columnwidth]{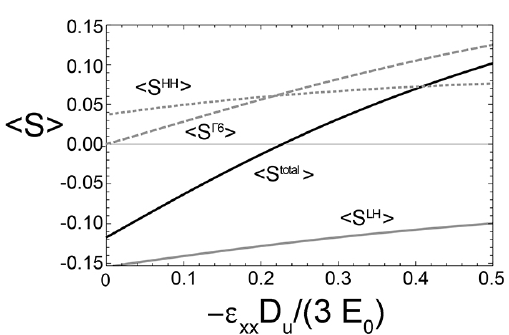}
\caption{(color online) Orbital resolved (grey lines) and total spin (black line) of the TSS with positive pseudospin (mirror eigenvalue $+i$) for the $(010)$ [$(001)$] plane at the mirror invariant line $k_z \equiv 0$ [$k_y \equiv 0$]. For small strain, the physical spin is antiparallel to the pseudospin, while for large strain they are parallel.}
\label{fig:fig4}
\end{figure} 
To gain more insight into this spin texture topological phase transition, we have computed the orbital resolved projected spin eigenvalue of the TSS along the mirror invariant lines $k_{y,z} \equiv 0$ for the $(001)$ and $(010)$ surface terminations respectively, by varying the strain magnitude [c.f. Fig.~\ref{fig:fig4}]. The TSS with positive pseudospin, {\it i.e.} mirror parity eigenvalue $+i$, is an admixture of $\ket{J,J_{y,z}}= \ket{1/2,1/2}$ $\Gamma_6$, $\ket{3/2,3/2}$ HH and $\ket{3/2,-1/2}$ LH states. For small strain, the TSS has a predominant LH orbital character which implies that the physical spin is antiparallel to the pseudospin. By continuously increasing the strain magnitude, the TSS starts to acquire a sizeable $\Gamma_6$ and HH character which ultimately reverse the spin direction to be parallel to the pseudospin.

\paragraph{Conclusions --} We have classified the topological surface state spin texture of bulk uniaxially strained HgTe using the close relations among spin texture  winding number, mirror Chern numbers and orbital character of the topological surface states. We have shown that assuming a strain along the $\hat{x}$ direction, the spin texture at the ${\mathcal C}_{4v}$ point-group symmetric $(100)$ plane exhibit a conventional left-handed helical structure, while considering the ${\mathcal C}_{2v}$ symmetric  $(010)$ and $(001)$ surface terminations the nature of the spin texture strongly depend on the orbital character of the topological surface states. The topological charge of the spin texture vortex structure can be indeed flipped by increasing the strain magnitude. This phenomenon is entirely due to the strain dependence of the orbital character of the TSS and occurs without any bulk bandgap closing-reopening point or change in the bulk MCN values, making such a topological phase transition very different from the ones proposed in for instance SmB$_6$ \cite{leg15} and  HgTe$_x$S$_{1-x}$ \cite{rau15}.

We acknowledge the financial support of the Future and Emerging Technologies (FET) programme within the Seventh Framework Programme for Research of the European Commission under FET-Open grant number: 618083 (CNTQC). This work has been supported by the Deutsche Forschungsgemeinschaft under Grant No. OR 404/1-1 and SFB 1143. JvdB acknowledges support from the Harvard-MIT Center for Ultracold Atoms.

\begin{appendix} 
\section{Kane model Hamiltonian}

As long as the intrinsic bulk-inversion asymmetry of the zinc-blende crystal structure is not taken into account, the six-band Kane model Hamiltonian reads
\begin{equation}
{\cal H}= \left( \begin{array}{ccc} {\cal H}^{6 , 6} & {\cal H}^{6 , 8} \\ 
  {\cal H}^{8 , 6} &  {\cal H}^{8 , 8} 
\end{array}
\right)
\label{eq:hamiltonian}
\end{equation}
with  the expression of the Hamiltonian subblocks ${\cal H}^{\alpha , \beta}$ listed in Table \ref{tab:table1}. They are expressed in terms 
of the usual Pauli matrices $\sigma_{x,y,z}$ , the 
$J=3/2$ angular momentum matrices $J_x = \sqrt{3} / 2 \,  {\cal I} \otimes \sigma_x + ( \sigma_x \otimes \sigma_x + \sigma_y \otimes \sigma_y ) / 2$, $J_y = \sqrt{3} / 2 \,  {\cal I} \otimes \sigma_y  + ( \sigma_y \otimes \sigma_x - \sigma_x \otimes \sigma_y ) / 2$, $J_z = \sigma_z \otimes {\cal I} + {\cal I} \otimes \sigma_z  / 2$ with ${\cal I}$ the identity matrix and the following  $T_i$ matrices:
\begin{eqnarray*} 
T_x& = &\dfrac{1}{3 \sqrt{2}} \left( \begin{array}{cccc}- \sqrt{3} & 0 & 1 & 0 \\ 0 & -1 & 0 & \sqrt{3} \end{array} \right) \\
T_y &= & \dfrac{-i}{3 \sqrt{2}} \left( \begin{array}{cccc} \sqrt{3} & 0 & 1 & 0 \\ 0 & 1 & 0 & \sqrt{3} \end{array} \right) \\
T_z &= &\dfrac{\sqrt{2}}{3} \left( \begin{array}{cccc} 0 & 1 & 0 & 0 \\ 0 & 0 & 1 & 0 \end{array} \right) \\
T_{xx}&=& \dfrac{1}{3 \sqrt{2}} \left( \begin{array}{cccc} 0 & -1 & 0 & \sqrt{3} \\ -\sqrt{3} & 0 & 1 & 0 \end{array} \right) \\
T_{yy}&=& \dfrac{1}{3 \sqrt{2}} \left( \begin{array}{cccc} 0 & -1 & 0 & -\sqrt{3} \\ \sqrt{3} & 0 & 1 & 0 \end{array} \right) \\
T_{zz} & =& \dfrac{\sqrt{2}}{3} \left( \begin{array}{cccc} 0 & 1 & 0 & 0 \\ 0 & 0 & -1 & 0 \end{array} \right) \\
T_{yz} &=& \dfrac{i}{2 \sqrt{6}} \left( \begin{array}{cccc} -1 & 0 & -\sqrt{3} & 0 \\ 0 & \sqrt{3}  & 0 & 1 \end{array} \right) \\
T_{zx} &=&  \dfrac{1}{2 \sqrt{6}} \left( \begin{array}{cccc} -1 & 0 & \sqrt{3} & 0 \\ 0 & \sqrt{3}  & 0 & -1 \end{array} \right) \\
T_{xy} &=&  \dfrac{i}{ \sqrt{6}} \left( \begin{array}{cccc} 0 & 0 & 0 & -1 \\ -1 & 0  & 0 & 0 \end{array} \right) \\
\end{eqnarray*}

The parameters $F, \gamma_1, \gamma_2, \gamma_3$ describe the coupling to remote bands and are, as well as $P, E_0$, material-specific parameters whose values are reported in Table \ref{tab:table1}
We have considered for simplicity the axial approximation $\overline{\gamma} = ( \gamma_2 + \gamma_3 ) / 2$ with the warping parameter $\mu= (\gamma_3-\gamma_2) / 2 \equiv 0$  in order to make the bulk band structure isotropic in the $k_{x,y}$ plane.

\begin{table}[tb]
\centering
 \caption{Expressions of the Kane model in the axial approximation. Here $\left\{A, B \right\}$ denotes the anticommutator for the $A,B$ operators,  $c.p.$ cyclic permutations of the preceding term and we defined $B= \hbar^2 / ( 2 m_0)$ with $m_0$ the free electron mass. We also list the band structure parameters for HgTe of Ref.\cite{nov05} .}
\begin{ruledtabular}
\begin{tabular}{ccc}
Hamiltonian blocks & & {\bf k $\cdot$ p} interactions \\
\hline
${\cal H}^{6 , 6}$ & & $E_0 + B (2F + 1 ) {\bf k}^2 $ \\ [1ex]
${\cal H}^{6 , 8}$ & & $\sqrt{3} P\,  {\bf T \, \cdot \, k}$ \\ [1ex]
& &  $-B \gamma_1 {\bf k}^2 + 2 B \overline{\gamma} \left[ \left(J_x^2 - \frac{J^2}{3} \right) k_x^2 + c.p. \right] $ \\ [-1ex]
\raisebox{1.5ex}{$ {\cal H}^{8 , 8} $} & & $ + B \overline{\gamma} \left[ \left\{ J_x , J_y \right\} \left\{k_x , k_y \right\} + c.p. \right] $ \\ [1ex]
${\cal H}^{7 , 7}$ & & $-\Delta_0 - B \gamma_{1} {\bf k}^2$ \\
\end{tabular}
\end{ruledtabular}
\vspace{.2cm}
\begin{ruledtabular}
\begin{tabular}{cccccc} 
$E_0$ & $F$ & $P^2/ B $ & $\gamma_1$ & $\gamma_2$ & $\gamma_3$ \\
\hline
-0.3 eV & 0 & 18.8 eV & 4.1 & 0.5 & 1.3 
\end{tabular}
\end{ruledtabular}
\label{tab:table1}
\end{table}
\begin{table}[tb]
\centering
 \caption{Strain-induced terms in the six-band Kane model Hamiltonian}
\begin{ruledtabular}
\begin{tabular}{ccc}
Hamiltonian blocks & &strain-induced interactions \\
\hline
${\cal H}^{6 , 6}$ & & $C_{1} Tr \epsilon$ \\ [1ex]
& & $D_d Tr \epsilon + \frac{2}{3} D_u \left[ \left(J_x^2 - \frac{1}{3} J^2 \right) \epsilon_{xx} + c.p. \right]  $\\ [-1ex]
\raisebox{1.5ex}{$ {\cal H}^{8 , 8} $}  & & $\frac{2}{3} D_u^{\prime} \left[ \left\{J_x , J_y \right\} \epsilon_{x y} + c.p. \right]$  \\[1ex]
\end{tabular}
\end{ruledtabular}
\label{tab:table2}
\end{table}
Effects of strain can be taken into consideration by applying the formalism of Bir and Pikus. They lead to additional terms in the eight-band Kane model Hamiltonian  proportional to the strain tensor ${\boldsymbol \epsilon}$ and expressed 
in terms of the  conduction and valence band deformation potentials  $C_{1}, D_u, D_u^{\prime}$  [see Table\ref{tab:table2}]. 

\begin{figure}
\includegraphics[width=.9\columnwidth]{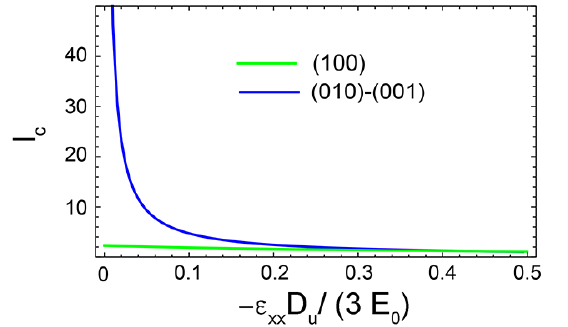}
\caption{(color online) Behavior of the topological surface state penetration depth in compressively strained HgTe as a function of the strain magnitude $\epsilon_{xx}$ at the $(100)$, $(010)$, and $(001)$ surfaces.}
\label{fig:figsupp1}
\end{figure} 

\section{Evaluation of the Mirror Chern numbers}
We recall that the (first) Chern number of a generic  multi-band system is defined as an integral over a pseudo-2-form
\begin{equation}
(\nu)_j=\frac{i}{2\pi}\int_{\cal M}d^2k\;\mbox{Tr}(P_j\wedge dP_j\wedge dP_j),
\label{eq:cherngeneral}
\end{equation}
where $j$ is the index of a non-degenerated band, $P_j=\ket{j}\bra{j}$ the related projector, $d$ is the exterior derivative, $\wedge$ is the wedge product and $\cal M$ is the momentum manifold, which corresponds to the one-point compactified infinite momentum plane $\mathbb{R}\cup\{\infty\}$ for the present long-wavelength continuum theory. In the presence of a full band gap in which the Fermi energy lies, the sum over all occupied bands $\sum_{j \in {\it occ}} \nu_{j}$ is a well defined integer topological quantity even in the presence of degeneracies of the bands over the manifold $\cal M$. 

At the mirror invariant planes $k_{x,y,z} \equiv 0$ the Kane model Hamiltonian commutes with the corresponding mirror operators. The corresponding Hamiltonians for the states with $+i$ mirror parity can be expanded in terms of the  $SU(3)$ matrices as
\begin{equation}
{\cal H}_{+i}=\varepsilon_0(\mathbf{k}_\parallel)\lambda_0+\sum_{i=1}^8\lambda_id_i(\mathbf{k}_\parallel), 
\end{equation}
where $\{\lambda_0,\lambda_1....\lambda_8\}$ are the $1+8$ Gell-Mann matrices with $\lambda_0$  the identity. In addition the $d$'s define an eight-dimensional vector in momentum space. The projector for each band can be represented by 
\begin{equation}
P_j=\frac{1}{3}(1+\sqrt{3}\mathbf{n}_j\cdot\boldsymbol{\lambda}),
\end{equation}
where $j$ enumerates the bands from $1$ to $3$. The appearing $\mathbf{n}_j$ defines a point on the unit sphere in an eight-dimensional real Euclidean space restricted by the condition $\mathbf{n}_j\ast\mathbf{n}_j=\mathbf{n}_j$ where $\ast$ is the $SU(3)$ star product. The vector $\mathbf{n}_j$ lives in a four dimensional orbit of a $S^7$ sphere which corresponds to an element of the complex projective plane $\mathbb{C}P^2$.  The relation between $\mathbf{n}_j$ and the normalized $\hat{\mathbf{d}}=\mathbf{d}/|\mathbf{d}|$ is found out to be
\begin{equation}
\mathbf{n}_{j}=\frac{1}{\gamma_{j}^2-1}(\gamma_{j}\hat{\mathbf{d}}+\hat{\mathbf{d}}\ast\hat{\mathbf{d}})
\end{equation}
where
\begin{equation}
\gamma_{j}=2\cos\left(\frac{1}{3}\mbox{arccos}(\hat{\mathbf{d}}\cdot\hat{\mathbf{d}}\ast\hat{\mathbf{d}})+\frac{2\pi}{3}j\right).
\end{equation} 
Eqs.(5),(6),(7)  provide an explicit expression for the mirror Chern numbers of the Kane model Hamiltonian.

\end{appendix} 


\begin{thebibliography}{9}
\bibitem{note}
In the remainder,  the effect of the bulk inversion asymmetry in the zincblende crystal structure of mercury chalcogenides will be considered as a small perturbation \cite{fu07}, and thus neglected.
\end{thebibliography}
\end{document}